\documentclass[11pt,letterpaper]{article}
\usepackage{jheppub}
\usepackage{amsmath,amssymb,amsfonts,bm}
\usepackage{graphicx,wrapfig}
\usepackage{multirow}
\usepackage{verbatim}
\usepackage{appendix}
\usepackage{rotating}
\usepackage{ytableau}
\ytableausetup{boxsize=.2cm}
\usepackage{cancel}

\numberwithin{equation}{section}

\newcommand{\CTb}{\mathcal{T}}
\newcommand{\CTT}{\mathcal{T}_{I\!I}}
\newcommand{\CBT}{\mathcal{B}_\mathcal{T}}

\newcommand{\bea}{\begin{eqnarray}}
\newcommand{\eea}{\end{eqnarray}}
\newcommand{\be}{\begin{equation}}
\newcommand{\ee}{\end{equation}}
\newcommand{\bse}{\begin{subequations}}
\newcommand{\ese}{\end{subequations}}

\newcommand{\wt}{\widetilde}

\newcommand{\ol}{\overline}

\newcommand{\ie}{\emph{i.e.}}
\newcommand{\cf}{\emph{cf.}}

\newcommand{\Z}{{\mathbb Z}}
\newcommand{\R}{{\mathbb R}}

\newcommand{\Tr}{{\rm Tr \,}}

\newcommand{\pd}{\partial}

\newcommand{\CB}{\mathcal{B}}

\newcommand{\CH}{\mathcal{H}}
\newcommand{\CI}{\mathcal{I}}

\newcommand{\CM}{\mathcal{M}}
\newcommand{\CN}{\mathcal{N}}

\newcommand{\CT}{\mathcal{T}}

\newcommand{\CZ}{\mathcal{Z}}

\title{An $E_7$ surprise}

\author[1]{Tudor Dimofte}
\author[2]{Davide Gaiotto}

\affiliation[1]{Institute for Advanced Study, Einstein Dr., Princeton, NJ 08540, USA}
\affiliation[2]{Perimeter Institute for Theoretical Physics, 
Waterloo, Ontario, Canada N2L 2Y5}

\abstract{We explore some curious implications of Seiberg duality for an $SU(2)$ four-dimensional gauge theory with eight chiral doublets. We argue that 
two copies of the theory can be deformed by an exactly marginal quartic superpotential so that they acquire an enhanced $E_7$ flavor symmetry. 
We argue that a single copy of the theory can be used to define an $E_7$-invariant superconformal boundary condition for a theory of $28$ five-dimensional free hypermultiplets.
Finally, we derive similar statements for three-dimensional gauge theories such as an $SU(2)$ gauge theory with six chiral doublets or $N_f=4$ SQED.}

\begin{document}

\maketitle

\section{Introduction}

The superconformal index  is a powerful, computable invariant of ${\cal N}=1$ superconformal field theories \cite{KMMR-index,Romelsberger-count}.%
\footnote{This quantity could more appropriately be called a ``sphere index'': it can be defined for non-conformal theories as well, 
as the Witten index of the Hilbert space of a supersymmetric compactification on $S^3$. This rigorously justifies the standard strategy of computing the index
directly from the UV description of a theory. We will often simply say ``sphere index'' in this note.} %
The comparison of indices computed from different UV descriptions of a theory has provided strong checks of old dualities, and predicted a few new ones. 
By a lucky accident, the special functions which enter the indices of ${\cal N}=1$ gauge theories have been subject of recent, independent mathematical investigation.
These results provide a toolbox of special function identities which can be used to test and predict field theory dualities. 

The authors of \cite{Spir-thetaintegrals, Rains-hypgeom, SV-E7} built a special function with a rich group of hidden symmetries, based on the Weyl group $W(E_7)$ of $E_7$. These symmetries control 
Seiberg-like dualities of a simple ${\cal N}=1$ theory:  $SU(2)$ $N_f=4$ SQCD \cite{Seiberg-duality, CSST-duals, IP-duals}.
We will denote this theory simply as ``$\CTb$'' in the following. 
In each dual frame one encounters the same theory $\CTb$, with different choices of flavor quantum numbers for the elementary fields
and additional superpotential couplings to gauge-invariant chiral fields.
If one ignores these extra chiral fields, the $W(E_7)$ duality group acts on the flavor fugacities in the index
exactly as if the manifest $SU(8)$ flavor symmetry of the theory was embedded in a larger $E_7$ flavor group. 

In this note, we attempt to tweak the physical setup in such a way that $E_7$ may become the actual flavor symmetry of a theory. The first step is to 
make sure $W(E_7)$ acts as a group of self-dualities of a single theory, rather than dualities between distinct theories. In other words,  
we need to eliminate the extra gauge-invariant fields that accompany the $W(E_7)$ duality transformations of $\CTb$. 
We can do so in two different ways: we can either couple together two copies of $\CTb$ to give a new SCFT $\CTT$, 
or we can couple $\CTb$ to a set of $28$ free five-dimensional hypermultiplets
to define a superconformal boundary condition $\CBT$.  

In either case, we arrive at a theory whose index has a manifest $E_7$ flavor symmetry, 
which extends the naive $SU(8)$ flavor symmetry inherited from $\CTb$.  By itself, this does not guarantee 
that the theory has a true $E_7$ flavor symmetry. Our theories  have a large space of exactly marginal deformations $\CM$ where the $SU(8)$ flavor symmetry is broken manifestly to a rank $7$ subgroup.
The $W(E_7)$ duality symmetry acts non-trivially on $\CM$, and predicts the existence of several loci $\CM_\pi$ where the flavor symmetry 
is enhanced back to a distinct $SU(8)_\pi$ flavor group. The index of the theory is independent of exactly marginal superpotential deformations, and thus will be invariant under all these enhanced $SU(8)_\pi$
flavor symmetries, which combine to the $E_7$ symmetry group we are after.  

It is natural to conjecture the existence of a very special point in moduli space, where the flavor symmetry is truly enhanced to $E_7$. Indeed, at any fixed point of $\CM$ under the action of 
a generator of the duality group the theory will automatically enjoy the $E_7$ flavor enhancement. We can subject this conjecture to a stringent test, 
based on the analysis of \cite{GKSTW}: as we move in $\CM$, marginal chiral operators can only appear or disappear from the spectrum if conserved currents with the same quantum numbers appear or 
disappear as well. We can start from a region of $\CM$ where we know the marginal chiral operator content of the theory, transport it to the conjectural $E_7$ point, and test if the 
operators organize themselves into irreducible representations of $E_7$. 

The four-dimensional SCFT $\CTT$ passes this test beautifully. The boundary condition $\CBT$ passes the test as well, and also satisfies stringent constraints 
which follow from the $U\!Sp(56)$ flavor symmetry of the bulk theory. Thus we conjecture that a point of enhanced $E_7$ flavor symmetry exists for both. 
The two conjectures are not independent: if the boundary condition $\CBT$ has enhanced flavor symmetry then $\CTT$ will have it as well, as it can be engineered 
from the compactification of the five-dimensional theory on a segment.%
\footnote{The authors of \cite{SV-E7} present an infinite family of theories $\CT_N$ that have the same $W(E_7)$ network of dualities, namely $U\!Sp(2N)$ gauge theories coupled to 
$8$ chirals in a fundamental representation, and one in the antisymmetric tensor representation. Although our construction can be applied to those theories as well, 
it is easy to argue that the final result is just $N$ copies of $\CTT$ or $\CBT$. }

We conclude the note with a construction that was the original motivation for this work. We compactify the $\CBT$ boundary condition on a circle, with a specific choice of flavor Wilson line, 
and arrive at an $SO(12)$-invariant boundary condition $\CBT^*$ for a theory of $16$ four-dimensional free hypermultiplets, defined either by coupling to a 3d $\CN=2$ $SU(2)$ gauge theory with six flavors,
or to $\CN_f=4$ SQED. This boundary condition plays an important role in an upcoming paper \cite{DGGV-hybrid} on duality walls in four-dimensional ${\cal N}=2$ gauge theories. The partition functions for the decoupled 3d boundary theories and their functional identities here have been examined recently in \cite{TV-6j}. The mathematical derivation of these identities given in  \cite{TV-6j}
was an important inspiration for this note.

We hope that our basic methods here will be useful to argue for enhanced symmetry in other contexts. An interesting and closely related example%
\footnote{We thank G. Vartanov and S. Razamat for pointing out this example.} %
is four-dimensional $\CN=2$ Seiberg-Witten theory with gauge group $SU(2)$ and four fundamental hypermultiplets. The index of this theory was found to have an $F_4$ symmetry that governs the structure of S-duality transformations  \cite{vdBult-F4, Rastelli-2dQFT}. We leave it to the curious reader to find a modification of the theory 
for which enhanced $F_4$ symmetry appears at some point in the parameter space.


\section{A review of $SU(2)$ $N_f=4$ SQCD}

In this section we review some of the properties of our main actor, the theory $\CTb$.
This SCFT  is defined in the UV as an $SU(2)$ gauge theory coupled to eight chiral fields $q_\alpha^i$, transforming as a doublet of the $SU(2)$
gauge group, and a fundamental representation of the $SU(8)$ flavor group. The non-anomalous R-symmetry in the IR is a linear combination of the UV R-symmetry and the 
$U(1)$ axial symmetry rotating all quarks in the same way.

In the absence of a superpotential, the $SU(8)$ flavor group is unbroken, and the gauge coupling flows to a strongly-coupled fixed point in the IR, 
until the gauge-invariant chiral quark bilinears 
\begin{equation}
M^{ij} \,=\, q^i q^j\, \equiv\,  \epsilon^{\alpha \beta} q_\alpha^i q_\beta^j
\end{equation}
acquire dimension $3/2$, matching the IR R-charge $1$.
The quartic chiral operators
\begin{equation} \label{Oquartic}
O^{ij;kt} = M^{ij} M^{kt}
\end{equation}
are thus marginal. Naively, one might expect $\frac12 28(28+1)=406$ operators of this form, but in fact all the $O^{ij;kt}$ are not independent, because $M^{ij}$ is a rank-two quark bilinear. The totally antisymmetric rank-four tensor of $SU(8)$ (of dimension 70) is ``missing'' from \eqref{Oquartic}, and we find that the $O^{ij;kt}$ transform in a single irreducible representations of $SU(8)$ of dimension $336$, labelled by the Young diagram $\raisebox{.1cm}{\ydiagram{2,2}}$\,.

\subsection{Seiberg duality}
Seiberg duality provides a dual description of the theory $\CTb$ \cite{Seiberg-duality}. We are instructed to take another copy of 
$\CTb$, which we can denote as ${}^S \CTb$, and to split the eight doublets into four ``quarks'' $u_a$ transforming in the fundamental representation of an $SU(4)_L$ flavor group, and 
four ``antiquarks'' $v_b$ transforming in the conjugate representation of an $SU(4)_R$. We then add a $SU(4)_L \times SU(4)_R \times U(1)_V$ 
invariant coupling to a set of $16$ gauge-invariant chiral fields $m^{ab}$,
\begin{equation}
W = m^{ab} u_a v_b\,.
\end{equation}
This is a relevant deformation, and it pushes the theory to a new IR fixed point where $m^{ab}$ has R-charge $1$. 
We will denote this operation on ${}^S \CTb$ as ``flipping'' the meson operators of ${}^S \CTb$.

This fixed point is conjectured to give a dual description of $\CTb$. The $SU(8)$ flavor symmetry 
of $\CTb$ is thus expected to be recovered in the IR as an enhancement of $SU(4)_L \times SU(4)_R \times U(1)_V$.
It is important to observe that the enhancement is non-trivial. The doublets of $\CTb$ would be denoted as $q^a$ and $q^{b+4}$, 
with the same $U(1)_V$ charge but conjugate $SU(4)_{L,R}$ representations with respect to the $u_a$, $v_b$ of ${}^S \CTb$. 
At the level of gauge-invariant operators, this works because baryons $u_au_b$ and $v_av_b$ transform as 
antisymmetric tensors of $SU(4)_{L,R}$, which are self-conjugate, and can match the baryons of $\CTb$ as
\begin{equation}
M^{ab} = \epsilon^{abcd} u_c u_d \qquad \qquad M^{a+4,b+4} =  \epsilon^{abcd} v_c v_d .
\end{equation} The mesons $M^{a,b+4}$ of $\CTb$
match the $m^{ab}$ operator, while the mesons of ${}^S \CTb$ are killed by the superpotential. 

As an exercise in preparation for the rest of the paper, it is useful to deform the theory~$\CTb$ by a superpotential containing the square of the mesons,
\begin{equation} \label{Mdef}
\sum_{1 \leq i,j \leq 4} M^{i,j+4} M^{j,i+4}\,.
\end{equation}
This is an exactly marginal deformation, which gives rise to a line of fixed points with 
$SU(4)_d \times U(1)_V$ flavor symmetry. 

In the dual description, the superpotential term gives a mass to $m_{ab}$, which can be integrated away, 
leaving behind a quartic superpotential in the quarks of the same form as \eqref{Mdef}, but roughly inverse coefficient. The flavor symmetry is broken to $SU(4)_d\times U(1)_V$ where $SU(4)_d$ is the diagonal of $SU(4)_L$ and $SU(4)_R$.
Therefore, Seiberg duality maps the line of fixed points back to itself. This implies that there should be two distinct 
loci where the $SU(4)_d \times U(1)_V$ flavor symmetry is enhanced to $SU(8)$ in two different ways, and we recover 
either $\CTb$ or ${}^S \CTb$. 

It is important to note that this picture makes sense only because we can keep track of the flavor information along the fixed line. 
If we ignore the flavor symmetry data, we should think in terms of a $\Z_2$-quotient of the moduli space, with a single locus of enhanced symmetry and 
an orbifold point. In the following, we will always try to keep track of flavor symmetry data, and thus we will only quotient exactly marginal moduli spaces 
by dualities that commute with the flavor group.

The match of marginal operators along this line of fixed points is already rather non-trivial. 
At the $\CTb$ and ${}^S \CTb$ loci we have $336$ operators in an irreducible representation of the 
$SU(8)$ and ${}^S SU(8)$ enhanced flavor symmetries. We can decompose them into irreducible representations of $SU(4)_L\times SU(4)_R\times U(1)_V$ and then into representations of the actual broken flavor group $SU(4)_d\times U(1)_V$. Schematically,
\begin{align}
 336 &\to (6,6)_0+(10,\ol{10})_0 + (20',1)_4+(1,20')_{-4}+(\ol{20},\ol{4})_2+(4,20)_{-2} \label{336decomp}\\
  &\hspace{-.3cm}\to (1+15+20')_0+(1+15+84)_0+(20')_4+(20')_{-4}+(6+10+64)_{2}+(6+\ol{10}+64)_{-2}\,, \notag
\end{align}
where the subscript denotes the $U(1)_V$ ``baryon number'' charge. Thus, the $(6,6)_0$ representation here corresponds to a product of baryons and anti-baryons, the $(10,\ol{10})_0$ to meson bilinears, the $(20')_{\pm 4}$'s to products of two baryons or two anti-baryons, etc. (The $20'$ denotes the self-conjugate representation of $SU(4)$ with Young diagram $\raisebox{.1cm}{\ydiagram{2,2}}$\,.)

The flavor assignment to elementary fields in $\CTb$ and ${}^S \CTb$ differ by the sign of $U(1)_V$ charges.
Operators made only of baryons and anti-baryons of $\CTb$ are not lifted away from the special points, and 
match nicely with the corresponding operators at the ${}^S \CTb$ locus. 
The meson bilinears $(1+15+84)_0$ have no $U(1)_V$ charge, and thus match trivially, though those transforming in the $15$ are lifted by coupling to the fifteen broken currents of $SU(4)_L \times SU(4)_R$ away from the enhanced loci. One can demonstrate this more explicitly by observing that the lifted operators are precisely those obtained by acting with broken flavor currents on the deformation \eqref{Mdef}.

The products of a meson and a baryon $(6+10+64)_{2}$ are more interesting. They contain a single $SU(4)_d$ representation which does not appear among 
the $(6+\ol{10}+64)_{2}$ operators in ${}^S \CTb$ with the same $U(1)_V$ charge:
a symmetric fundamental tensor $10$ of $SU(4)_d$. The products of a meson and an anti-baryon contain a symmetric anti-fundamental tensor $\ol{10}$. 
Together with the two anti-symmetric tensors $(6)_{\pm 2}$, this is exactly the representation content of the broken block off-diagonal currents in $SU(8)$. 
These operators are thus lifted 
when we move away from the $\CTb$ point. All other marginal operators survive along the line of fixed points, and match 
the operators at the ${}^S \CTb$ locus.

\subsection{Seiberg-like dualities}
\label{sec:Slike}

There is a second dual description of $\CTb$, introduced by Cs\'aki \emph{et al.} \cite{CSST-duals}. It also involves a copy of $\CTb$,
which we can denote as ${}^C \CTb$, deformed by coupling the baryons and anti-baryons to gauge-invariant chiral fields 
in the antisymmetric tensors of $SU(4)_L$ and $SU(4)_R$ respectively, \ie\ by flipping the baryon and anti-baryon operators 
of ${}^C \CTb$. Again, this breaks the ${}^C SU(8)$ flavor symmetry to $SU(4)_L \times SU(4)_R \times U(1)_V$, enhanced to
the $SU(8)$ of $\CTb$ in the IR. 

We can repeat the analysis above, considering a second line of fixed points with $SU(4)_d \times U(1)_V$ symmetry 
that arises  from the deformation of the $\CTb$ theory by the quartic operator 
built pairing up baryons and anti-baryons. In the dual description, this gives masses to the gauge-invariant chiral fields, and leaves behind 
a superpotential of the same form. Thus we expect two loci of $SU(8)$ enhancement along this line of fixed points, where either $\CTb$ or ${}^C \CTb$
emerge. 

The combination of the two dualities above gives a third dual description, introduced by Intriligator and Pouliot \cite{IP-duals},
where one flips all the $\wt M_{ij}$ bilinear operators of a copy ${}^{IP} \CTb$ of $\CTb$ in which the chiral matter is taken to transform in an anti-fundamental
representation of $SU(8)$. The deformation preserves $SU(8)$, and flows in the IR to $\CTb$, with the chiral matter in the fundamental of $SU(8)$. 

We can consider the most general $SU(4)_d \times U(1)_V$ invariant deformation of $\CTb$ by a quartic superpotential. Inside the $336$ of $SU(8)$ we find exactly two independent
$SU(4)_d \times U(1)_V$ invariant operators --- 
the two singlets in the second line of \eqref{336decomp} --- which lead to a (complex) two-dimensional space of exactly marginal deformations. We should find 
$\CTb$, ${}^S \CTb$,${}^C \CTb$ and ${}^{IP} \CTb$ as four loci where the flavor symmetry is enhanced to $SU(8)$ in various ways.

\subsection{The $W(E_7)$ duality network}
\label{sec:E7net}

In defining Seiberg duality, we make an arbitrary choice of a splitting of $8$ doublets into quarks and anti-quarks, 
corresponding to a choice of $SU(4)_d\times U(1)_V$ inside $SU(8)$. 
Let us focus on a set of choices related by all possible permutations of the $8$ chiral multiplets, rather than 
more general flavor rotations. This insures that a common Cartan sub-algebra of $SU(8)$ remains manifest in all dual frames,
and allows a meaningful comparison among them.  
 This leads to $35$ distinct Seiberg duality operations, and $35$ Cs\'aki-et-al
dualities, where we take a copy of $\CTb$ and flip an appropriate subset of $16$ mesons, or $12$ baryons and anti-baryons. 

A key observation of \cite{SV-E7} is that all these dualities do not compose to a larger set. Rather, 
their action closes on the set of $72$ dual descriptions of $\CTb$\,: the original, the $35+35$ Seiberg and Cs\'aki-et-al duals, and 
the Intriligator-Pouliot dual frames. The authors of \cite{SV-E7} (see also \cite{Khmelnitsky-E7}) point out a beautiful group-theoretic way to 
organize this network of dualities. The flavor charges of the elementary matter fields in different duality frames are related by 
reflections inside the common Cartan subalgebra. These reflections combine with the Weyl group $W(A_7)$ of $SU(8)$ to give the Weyl group of $E_7$. The 72 dual theories are labelled by elements of the coset $W(E_7)/W(A_7)$.

We can embed $SU(8)$ inside a fictitious $E_7$ group, so that the reflections literally coincide with the Weyl reflections of $E_7$. 
Notably, the $56$-dimensional vector representation of $E_7$ decomposes as $28 + \ol{28}$ under $SU(8)$, 
and each reflection $\pi$ in $W(E_7)$ either inverts or leaves invariant the 
charge vectors of the $28$ gauge-invariant bilinear operators of the theory. 
The corresponding duality operation on $\CTb$ flips the operators whose charges are inverted. 

A crucial fact that makes this duality structure possible is the observation that flipping the same operator twice 
is the same as doing nothing. Let us see this in detail. We start from a theory $T$ and an operator $O$, 
add a new chiral field $\phi$ and superpotential coupling
\begin{equation}
W = \phi O
\end{equation}
and flow to the IR to define a new theory $T'$ and operator $O' = \phi$. We repeat the transformation on $T'$ and $O'$, add a new chiral field $\phi'$ and 
superpotential coupling 
\begin{equation}
W = \phi' O'
\end{equation}
and flow to the IR. 
The second coupling gives a mass to $\phi$ and $\phi'$, allowing us to integrate them away. Thus we flow back to the original theory, 
and $\phi'=O$.

Now we can see why no new dual descriptions are generated by iterating Seiberg dualities and Cs\'aki-et-al dualities. 
Each duality flips the operators whose charges are inverted by the corresponding element $\pi$ of $W(E_7)$. 
The composition of two dualities $\pi$ and $\pi'$ flips some operators twice. If we remove these physically trivial double flips, 
we clearly arrive at the dual frame associated to the composition $\pi \circ \pi'$.


\section{The doubled theory $\CTT$}
\label{sec:doubled}

Consider now the product of two $\CTb$ theories. We will use conventions where 
one theory has $SU(8)$ fundamental chiral doublets, and the other $\wt {SU(8)}$ anti-fundamental chiral doublets.
We can denote this as 
\begin{equation}
\CTb \times \wt \CTb\,.
\end{equation}

The product theory can be deformed by an exactly marginal operator of the form 
\begin{equation} \label{WTT}
W = c \sum_{i,j} M^{ij} \wt M_{ij}\,
\end{equation}
that preserves a diagonal $SU(8)_d$ flavor symmetry inside $SU(8) \times \wt{SU(8)}$.
The marginal operator with this property is unique, since there is a unique singlet in the product $28 \times \ol{28}$. 
Thus we have a line of fixed points with $SU(8)_d$ flavor symmetry. 

As anticipated in the introduction, the index of the deformed product theory automatically enjoys a discrete $W(E_7)$ symmetry. We will see this explicitly in Section \ref{sec:ind-double}. Presently, however, we want to demonstrate that an $E_7$ flavor symmetry actually 
appears as an enhanced physical flavor symmetry at a special point $p^*$ on the line of $SU(8)_d$\,-invariant fixed points. 

Seiberg-like dualities act on the deformed theory in a rather simple way. 
For example, if we apply Intriligator-Pouliot duality on both factors of the product theory, 
the exactly marginal deformation becomes a mass term for the gauge-invariant fields 
in the dual description. If we integrate them away, we get back the superpotential deformation of the 
product theory $\wt \CTb \times \CTb$, with a coefficient that is roughly $1/c$.
Thus the doubled Intriligator-Pouliot duality maps the line of fixed points back to itself in a non-trivial way. 

It is also interesting to consider an Intriligator-Pouliot duality 
acting on the second factor of the product theory only. Then we get the product of two $\CTb$ theories, 
with superpotential of the form 
\begin{equation}
W = c M^{ij} m_{ij} + c' m_{ij} N^{ij}\,,
\end{equation}
where $N_{ij}$ are the bilinear operators of the second theory. The RG flow will change the 
coefficients of the superpotential until $m_{ij}$ has the correct dimension $3/2$. The line of exactly marginal deformations of the product theory 
sits somewhere in the $(c,c')$ plane, and Intriligator-Pouliot duality on both factors should correspond to the obvious $\Z_2$ symmetry that exchanges $c$ and $c'$.

It is even more useful to apply Seiberg duality to both theories. After we integrate away the gauge-invariant matter, 
we produce a new quartic coupling, involving the mesons of the two theories. The term coupling the baryons and anti-baryons of the two theories 
is unaffected. Thus the doubled Seiberg duality maps the line of $SU(8)_d$\,-invariant fixed points to a new line of fixed points, in the general space of 
$SU(4)_L \times SU(4)_R \times U(1)$\,-invariant fixed points produced by introducing independent couplings between the mesons, 
baryons and anti-baryons of the ${}^S \CTb \times {}^S \wt \CTb$ product theory:
\begin{equation}
b u_a u_b \tilde u^a \tilde u^b + d u_a v_b \tilde u^a\tilde v^b + b v_a v_b \tilde v^a \tilde v^b\,.
\end{equation}
The couplings $(b, d)$ parametrize a two-dimensional space of exactly marginal deformations.%
\footnote{Although we could naively deform the theory further 
by coupling baryons and anti-baryons differently, the operator $u_a u_b \tilde u^a \tilde u^b - v_a v_b \tilde v^a \tilde v^b$ is not exactly marginal. It is lifted by pairing up with 
a broken $SU(8) \times \wt{SU(8)}$ flavor current with the same quantum numbers.}

The $SU(8)_d$\,-invariant line lies somewhere in this two-dimensional space. Points on the line corresponding to small $c$ in the original theory will map to small $b$ and large $d$. 
Points at large $c$ will map to large $b$ and small $d$.

It is reasonable to conjecture that we can make the meson and baryon couplings of the dual theory the same (\ie\ $b=d$) by tuning $c$ to an intermediate point $p^*$ on the fixed line, and that this is the same point that is mapped to itself under the doubled Intriligator-Pouliot duality. If this conjecture holds true, then 
the theory at the fixed point $p^*$ has a very special property: it enjoys both the 
standard $SU(8)_d$ flavor symmetry from the description as a deformation of $\CTb \times \wt \CTb$, 
and the enhanced ${}^S SU(8)_d$ flavor symmetry from the description as a deformation of ${}^S \CTb \times {}^S \wt \CTb$.

The two $SU(8)_d$ symmetries share a common $SU(4)_L \times SU(4)_R \times U(1)$ subgroup, and combine necessarily 
into an $E_7$ enhanced flavor symmetry.
We denote the deformed product theory at this conjectural, special point $p^*$ in moduli space fixed by all doubled Seiberg-like dualities 
as $\CTT$. 

In order to test the conjecture that such a point exists, we can look at the chiral operators of $\CTT$. 
First, the bilinear operators $M^{ij}$ and $\wt M_{ij}$ are mixed by the double Seiberg-dualities, and 
combine neatly into a single $56$-dimensional fundamental representation of $E_7$. 

More interestingly, we can look at at the quartic marginal operators. Naively, one would deduce that the quartic operators of the form $MM$, $M \wt M$, $\wt M \wt M$ have the correct quantum numbers to fit into a
a symmetric tensor of $E_7$. This is not quite right, for a good reason. First, we know that the $MM$ operators are not all independent, because $M$ is a rank-$2$ quark bilinear, \cf\ \eqref{Oquartic}. They sit in the 336-dimensional representation of $SU(8)_d$. Similarly, the $\wt M\wt M$ operators sit in the $\ol{336}$. The $28^2=784$ operators of the form $M\wt M$ are independent, but $63$ of them (an adjoint of $SU(8)_d$) are lifted on the line of $SU(8)_d$\,-invariant fixed points by combining with the broken off-diagonal currents of $SU(8)\times \wt{SU(8)}$. Indeed, the product $28\times \ol{28}$ decomposes into irreducibles as
\be 28\times\ol{28} \,=\, 1+63+720\,, \label{2828} \ee
and we can easily identify the lifted $63$.

All in all, we find $336+336+1+720=1393$ marginal operators at a generic point on the $SU(8)_d$\,-invariant fixed line. At the conjectural point $p^*$ of enhanced $E_7$ flavor symmetry, 70 new marginal operators must also appear, as partners of the new flavor currents. They form a rank-four fully antisymmetric tensor of $SU(8)_d$.

This fits perfectly well. The symmetric tensor of $E_7$ is reducible, decomposing into a 133-dimensional adjoint representation and another 1463-dimensional irreducible representation. The marginal operators at the point $p^*$ compose beautifully into the $1463$ of $E_7$. They are the marginal operators of $\CTT$. Moreover, there is a unique singlet of $SU(8)_d$ --- evident in \eqref{2828}, and corresponding to the superpotential \eqref{WTT} --- that can deform $\CTT$ back to the $SU(8)_d$\,-invariant line of fixed points.

Thus we have a self-consistent conjecture: there is a special choice of superpotential \eqref{WTT} for which the doubled theory $\CTb\times \wt\CTb$ 
has enhanced $E_7$ flavor symmetry.


\section{The boundary theory}
\label{sec:bdy}

There is a second, less familiar, construction that can be used to remove the gauge-independent fields from the duality, or produce a flip-independent 
physical setup. This was introduced in \cite{DGG, DGG-index} for 3d $\CN=2$ SCFT's. The generalization to four dimensions is straightforward. 
The basic claim is that two four-dimensional $\CN=1$ SCFTs related by the flip of one operator 
can be associated to a single superconformal boundary condition for a five-dimensional free field theory, 
which consists of one hypermultiplet with the same quantum numbers as the operator. 

A hypermultiplet consists of four real scalar fields and a fermion. Hypermultiplets can be defined in up to six dimensions. 
The $SO(4)$ symmetry that rotates the four scalar fields into each other is the product of an $SU(2)_R$ R-symmetry and 
an $SU(2)_f$ flavor symmetry. More generally, for a set of $n$ free hypermultiplets 
we can identify an $USp(2n)$ flavor symmetry that commutes with the supercharges. 
A theory of free hypermultiplets can be seen, trivially, as a superconformal field theory. 
The superconformal group in five dimensions is (a non-compact real form of) the exceptional supergroup $F(4)$. It has
bosonic subgroup $SO(5,2)\times SU(2)_R$. A free hypermultiplet in five dimensions naturally has scaling dimension 
$3/2$. 

We are interested in BPS, superconformal boundary conditions for a theory of hypermultiplets. We specialize to five dimensions, but most of what we say 
generalizes to six, four or fewer dimensions in a straightforward way. A superconformal boundary condition should preserve a four-dimensional superconformal 
group, \ie\ an $SU(2,2|1)$ subgroup of $F(4)$. The four-dimensional $U(1)_R$ R-symmetry is the Cartan generator in $SU(2)_R$.
If we normalize it in a standard way, so that supercharges have charge $\pm 1$, the hypermultiplets have four-dimensional R-charge $\pm 1$
and dimension $3/2$. Indeed, they decompose under four-dimensional supersymmetry into an $SU(2)_f$ doublet $(X,Y)$ of complex chiral fields of R-charge $1$. 

Looking at the supermultiplet structure a bit more closely, it is easy to see that the derivative $\partial_n \ol Y$ in the direction normal to the boundary 
is the F-term of the chiral multiplet of which $X$ is the bottom component. A similar observation holds for $\partial_n \ol X$. This can be understood, say, 
by writing the 5d Lagrangian for the bulk theory in four-dimensional terms \cite{dWFO}. Then the Lagrangian includes a superpotential of the rough form 
\begin{equation}
\int dx_n d^4x \,d^2\theta\, Y \partial_n X\,.
\end{equation}

One additional consideration is in order. One may wonder what the behavior of the $X$ and $Y$ 
fields will be at a generic superconformal boundary condition $B$. In a general 
CFT, we are used to having bulk-to-boundary OPEs, where a bulk operator 
is expanded as a sum of boundary operators weighed by powers of the distance $x_n$ from the boundary.
We can write down such a general OPE for $X$, 
\begin{equation}
X \sim \sum_i x_n^{\Delta_i-3/2} \left[O_i + x_n O^{(1)}_i + \cdots \right]
\end{equation} where the superscript indicates descendants of a primary under the 4d conformal group, 
and apply the equations of motion for $X$ to it,
\begin{equation}
\partial_n^2 X + \partial_\parallel^2 X=0\,. 
\end{equation}

As the derivatives $\partial_\parallel$ produce descendants, we must have $\partial_n^2 x_n^{\Delta_i-3/2} = 0$, i.e. 
\begin{equation}
X \sim \left[O_0 + x_n O^{(1)}_0 + \cdots \right] + x_n \left[O_1 + x_n O^{(1)}_1 + \cdots \right]
\end{equation}
The immediate result of the analysis is that $X$ must be regular at the boundary, and the only primaries that appear 
are the boundary values of $X$, and of the normal derivative of $X$; and the same is true for $Y$. 

The theory of a single free hypermultiplet has two basic free BPS boundary conditions.
A free boundary condition has to set half of the fermions to zero. By supersymmetry, 
this implies that half of the scalars will receive Dirichlet boundary conditions. 
With no loss of generality, we can give Dirichlet b.c. to $Y$. Then supersymmetry implies that 
$X$ will have Neumann b.c.:
\begin{equation} \label{BX}
B_X\,:\qquad \partial_n X|_\partial =0 \qquad Y|_\partial=0\,.
\end{equation}

This boundary condition preserves a $U(1)_f$ subgroup of the $SU(2)_f$ flavor symmetry. 
There is a second free boundary condition that preserves the same $U(1)_f$ subgroup as 
$B_X$: 
\begin{equation} \label{BY}
B_Y\,:\qquad X|_\partial=0 \qquad \partial_n Y|_\partial=0\,.
\end{equation}
A general free boundary condition is obtained 
from the basic $B_X$ b.c. by $SU(2)_f$ rotations. The infinitesimal rotations can be implemented by adding a marginal boundary superpotential 
\begin{equation}
c_+ \int_\partial d^4x \,d^2\theta\, X^2\,.
\end{equation}

In general, if a boundary condition breaks a flavor symmetry, we must have an operator on the boundary that 
is the boundary value of the normal component of the corresponding bulk current (which has a non-singular bulk-to-boundary OPE because of the bulk conservation law). 
For a free theory of bulk hypermultiplets, 
written as $2n$ chiral fields $Z^A$ in a fundamental representation of $USp(2n)$, the normal components of the flavor currents take the form (we denote the symplectic form as $\omega^{AB}$)
\begin{equation}
 Z^{(A} \omega^{B)C}\overleftrightarrow \partial_n \bar Z_C = \int d^2 \theta \,Z^A Z^B + c.c.\,,
\end{equation}
and are descendants of the marginal chiral operators $Z^A Z^B$ and/or its complex conjugate.
As the dimension of bulk currents are fixed and the normal component of bulk currents has a good limit at the boundary, 
the bilinears $Z^A Z^B$ will not diverge at the boundary. If the corresponding symmetry is unbroken they will go to zero at the boundary, 
otherwise they will define protected chiral or anti-chiral boundary operators. 
For example, free boundary conditions split the $Z^A$ into two sets $(X^i, Y_i)$, with  Neumann and Dirichlet b.c. and break $U\!Sp(2n)$ to $U(n)$. The 
broken symmetries are associated to the bilinears $X^i X^j$ at the boundary and their complex conjugates. 

In general, we should write 
\begin{equation}
\left(Z^A Z^B\right)\big|_\partial = c_a^{AB} O^a
\end{equation}
for some canonically normalized boundary operators $O^a$. The coefficients $c_a^{AB}$ are functions of the couplings, and the rank of the $c_a^{AB}$ matrix will drop
at loci where a larger amount of bulk flavor symmetry is preserved 
by the boundary condition; however, no marginal boundary operators will appear or disappear. This is an important difference from the purely four-dimensional 
flavor symmetry enhancement we encountered in previous sections. 

We can define a larger class of 
boundary conditions for a free hypermultiplet if we add some extra four-dimensional degrees of freedom at the boundary, in the form of some 4d SCFT $T$ with a chiral operator $O$.
We can couple it to the 
five-dimensional free theory with $B_X$ b.c. by a superpotential 
\begin{equation}
W = O X|_\partial\,.
\end{equation}
Such a superpotential coupling modifies the Dirichlet b.c for $Y$ to $Y=O$, and the other boundary conditions accordingly. 
If we flow to the IR, we expect this construction to define a new superconformal boundary condition $B_X[T]$ for the theory of a bulk free hypermultiplet. 
We can repeat the same exercise starting from the b.c. where $X$ has Dirichlet b.c., and couple the boundary value of $Y$ 
to a different theory $T'$ with an operator $O'$. This will define in the IR a boundary condition $B_Y[T']$.

A very special case is when $T$ is a theory of a free chiral multiplet $\phi$. 
Then the superpotential 
\begin{equation}
W = \phi X|_\partial
\end{equation}
enforces Dirichlet b.c. for $X$, while $Y=\phi$ just means that the Dirichlet b.c. for $Y$ has been relaxed. 
Thus in the IR we recover the opposite boundary condition \eqref{BY}. In other words, 
$B_X[\phi] = B_Y[\emptyset]$. 

This is an example of a more general phenomenon: given any BPS boundary condition $B$, there is a standard strategy 
to produce two 4d SCFTs $T$ and $T'$ such that $B = B_X[T] = B_Y[T']$, detailed in \cite{DGG, DGG-index} and similar to constructions in \cite{DGG-defects}.
The theories $T$ and $T'$ can be extracted by placing the 5d theory on a segment, with the $B$ boundary condition on one end and 
the appropriate Neumann/Dirichlet b.c. on the other end, and flowing to the IR. 
For example, if we have Dirichlet b.c. on $X$ at the other end, the resulting theory $T$ will have a distinguished operator $O$ coinciding with the 
boundary value of $Y$ at that end. This operator $O$ is used to produce the $B_X[T]$ description of the boundary condition. 
A corollary of this construction is that if we have two 4d SCFTs $T$ and $T'$ related by a flip, we should naturally think about a single boundary condition 
$B = B_X[T] = B_Y[T']$ for an appropriate free 5d theory. We will apply this strategy momentarily to the dual descriptions of $SU(2)$ $N_f=4$. 

\subsection{A boundary condition $\CB_\CT$ for $28$ hypermultiplets}
In order to be able to ignore the flip of any bilinear operator $M_{ij}$ of $\CTb$, we should couple it to a theory of $28$ 5d hypermultiplets. The bulk theory has $U\!Sp(56)$ flavor symmetry, and the hypermultiplets can be split into chirals $X_{ij}$ and $Y^{ij}$ that transform as an antisymmetric tensor and a conjugate antisymmetric tensor, respectively, of a certain $SU(8)$ subgroup of $U\!Sp(56)$. We then couple to $\CTb$ via an $SU(8)$-invariant superpotential
\begin{equation}
W = c \sum_{i,j} M^{ij} X_{ij}\,. \label{MX}
\end{equation}

The first important observation is that this coupling is marginal. It is clearly exactly marginal, for the same reason for which the quartic superpotential deformations of the previous sections were exactly marginal: we can always adjust the gauge coupling as a function of $c$ so that $M^{ij}$ remains of dimension $3/2$, while the dimension of the 5d field $X_{ij}$ is fixed. 
Thus we have a fixed line of $SU(8)$-invariant superconformal boundary conditions. 

We can apply any of the Seiberg-like dualities to the system. If we apply Intriligator-Pouliot duality, we get a coupling of the $X_{ij}$ to a completely flipped 
${}^{IP} \CTb$ theory. This is the same as coupling of the $Y^{ij}$ directly to the ${}^{IP} \CTb$ theory (with no flips), with a coupling constant that is roughly the inverse of $c$. 
Thus the duality maps the $SU(8)$-invariant fixed line back to itself. 

If we apply Seiberg duality, we get a coupling of the $X_{ij}$ to 
the theory obtained from ${}^S \CTb$ by flipping the mesons. We can rewrite that as the coupling of ${}^S \CTb$
to a different half of the bulk hypermultiplets, consisting of $X_{aa'}$, $Y^{ab}$ and $X_{b b'}$, where $a,a'$ run from $1$ to $4$, 
and $b,b'$ from $5$ to $8$. The coefficients of the ``baryon'' $X$  superpotential couplings and the ``meson'' $Y$ superpotential couplings are in general different: the dual image of the original 
$SU(8)$ invariant fixed line is some line in a larger (three-dimensional) moduli space of $SU(4)_L \times SU(4)_R \times U(1)_V$-invariant boundary conditions. 

Following a reasoning that is entirely analogous to the one we gave for the existence of $\CTT$, we can argue that as we increase $c$ from the 
value where the $\CTb$ degrees of freedom almost decouple from the bulk, to the value where the ${}^{IP} \CTb$ degrees of freedom almost decouple from the bulk,
we should hit an intermediate self-dual point $p^*$ where the coefficients of the meson and baryon couplings in the Seiberg-dual description become equal, and thus the flavor symmetry is enhanced to an ${}^S SU(8)$ flavor symmetry that combines with the original $SU(8)$ to give $E_7$. We denote the boundary condition at the self-dual point as $\CBT$. 

Now the $E_7$ flavor symmetry rotates the $X_{ij}$ and $Y^{ij}$ into each other, \ie\ it is a subgroup of the bulk flavor symmetry. 
Luckily, the fundamental $56$-dimensional representation of $E_7$ is pseudo-real, so we can organize the bulk hypermultiplets 
into a single half-hypermultiplet in the fundamental of $E_7$. 

The comparison of marginal operators from the decoupling region $c \to 0$ to the conjectural $E_7$ point works very well. 
The main difference from the case of $\CTT$ is the fact that the bilinears $X_{ij} X_{kt}$ include the totally antisymmetric $70$ representation of $SU(8)$ --- unlike the $MM$ operators of $\CTT$. 
These extra marginal operators are needed: while four-dimensional broken currents combine with marginal operators and lift them, 
five-dimensional currents broken at the boundary combine with boundary marginal operators, but do not lift them: 
the dimension of the 5d currents is fixed in the bulk. Thus the $70$ marginal operators play the role of boundary values for the $E_7$ currents broken along the line of exactly marginal deformations. 
The remaining marginal operators of $\CBT$, analogous to those of $\CTT$, sit in a dimension-$1463$ representation of $E_7$ and can neatly be associated to the 
$U\!Sp(56)$ bulk currents broken by the $E_7$-invariant boundary condition.  

Notice that the existence of $\CBT$ by itself would guarantee the existence of $\CTT$, as the latter can be engineered 
by putting the 5d theory on a segment, with  $\CBT$ boundary conditions at both ends.


\section{Indices}
\label{sec:indices}

The sphere index provides a beautiful way to illustrate and to verify the $E_7$ flavor symmetry of the theories discussed above. If a theory $\CT$ has flavor symmetry group $G$, then its index $\CI_\CT$ depends on a set of flavor fugacities $z_i$ ($i=1,...,{\rm rank}(G)$) corresponding to a maximal abelian subgroup of $G$. The index must then be invariant under the Weyl group $W(G)$, acting on the set of flavor fugacities. We will review the index of $\CTb$, following \cite{SV-E7}, and then construct the indices of the doubled theory $\CTT$ and the boundary condition $\CBT$, demonstrating that they are $W(E_7)$--invariant functions.

It is important to note that finding $W(G)$ symmetry in the index of a certain theory $T$ is a good indication but \emph{not} a guarantee that $\CT$ itself has flavor symmetry $G$. The index is invariant under superpotential deformations. Therefore, a theory $T$ may generically have marginal deformations --- invisible in the index --- that break its putative flavor symmetry to a subgroup of $G$. A careful analysis is then necessary to ensure that there exists a point in the marginal parameter space of $T$ with truly enhanced symmetry. We performed such an analysis above for $\CTT$ and $\CBT$.

\subsection{Index of $\CTb$ and its flips}

The sphere index for a 4d $\CN=1$ theory $T$ with $U(1)_R$ R-symmetry was defined in \cite{KMMR-index, Romelsberger-count, Romelsberger-calculate} (see also \cite{DolanOsborn}). It can be written in the form of a Witten index:
\be \label{4dindex}
\CI_\CT(z;x,y) = \Tr_{\CH(S^3)}(-1)^F e^{-\beta H} x^{2J_1+R}y^{2J_2}z^e\,,
\ee
where $R$ is the $U(1)_R$ R-charge of states on $S^3$; $J_1$ and $J_2$ are the angular momenta corresponding to Cartan generators of the rotation group $SO(4)\simeq SU(2)_1\times SU(2)_2$; and $e=(e_1,...,e_n)$ are the flavor charges corresponding to Cartan generators of the flavor symmetry group. The index is constructed with respect to a certain supercharge $Q$ in the supersymmetry algebra on $S^3\times \R$, and only gets contributions from states with $H = E-2J_1-\frac32 R = 0$, where $E$ is the energy.
It will be convenient for us to re-define the angular-momentum and R-charge fugacities in a more symmetric fashion as $x = \sqrt{pq},\, y=\sqrt{p/q}$. Then the index becomes
\be \label{4dpq}
 \CI_\CT(z;p,q) = \Tr_{\CH(S^3)}(-1)^F e^{-\beta H} p^{\tfrac R2+J_1+J_2}q^{\tfrac R2+J_1-J_2}z^e\,.
\ee

Crucially, the sphere index of a superconformal theory $T$ can be calculated using a weakly coupled UV Lagrangian theory $T_{UV}$ that flows to $T$ in the IR but need not itself be conformal. The index is invariant under RG flow. Thus, the index provides a perfect testing ground for Seiberg and Seiberg-like dualities that relate different UV descriptions of the same IR SCFT.

A simple but relevant fact is that a chiral multiplet with R-charge $r$ and flavor charge $f$ contributes to the index a factor
\be \label{Ichiral}
\CI_{\rm chiral}(z;p,q) = \Gamma((pq)^{\frac r2}z^f;p,q)\,,
\ee
where the \emph{elliptic Gamma function} is defined as
\be \label{G}
 \Gamma(z;p,q) = \prod_{m,n=0}^\infty \frac{1-p^{m+1}q^{n+1}z^{-1}}{1-p^mq^nz} \equiv \frac{(pqz^{-1};p,q)_\infty}{(z;p,q)_\infty}\,,
 \ee
with $(z;p,q)_\infty = \prod_{m,n=0}^\infty (1-p^mq^nz)$\,. The denominator of \eqref{Ichiral} is associated to modes of the scalar in a chiral multiplet, and the numerator comes from modes of a fermion in the conjugate anti-chiral.

Now, consider the theory $\CTb$: $SU(2)$ gauge theory with eight quark doublets. Let us give the fundamental chirals $q^i$ an R-charge $R(q^i)=1/2$, to match their superconformal R-charge assignment in the IR. Then the index can be computed as \cite{SV-E7}
\be \label{I1}
 \CI_\CT(z;p,q) = (p)_\infty(q)_\infty \oint \frac{ds}{4\pi is} \frac{\prod_{i=1}^8 \Gamma((pq)^{\frac14}z_is;p,q)\Gamma((pq)^{\frac14}z_is^{-1};p,q)}{
\Gamma(s^2;p,q)\Gamma(s^{-2};p,q)}\,,
\ee
where the prefactor contains $(p)_\infty(q)_\infty = \prod_{m=1}^\infty(1-p^m)(1-q^m)$. The eight flavor fugacities $z_i$ corresponding to the $SU(8)$ flavor group satisfy $\prod_{i=1}^8 z_i=1$. The numerator of the integrand clearly comes from the eight chiral doublets, while the denominator and the prefactors come from the $SU(2)$ gauge multiplet; the variable $s$ is the fugacity for the $SU(2)$ gauge symmetry, and the integral $\oint \frac{ds}{s}$ over the unit circle serves to project onto gauge-invariant states.
The function $\CI_\CT(z;p,q)$ is manifestly invariant under the Weyl group $W(SU(8))=S_8$, which permutes the $z_i$.

The various Seiberg-like dualities that generate the $W(E_7)$ duality network of $\CTb$ act in a very simple way on the index \eqref{I1}, as described in \cite{SV-E7}. Consider, for example, the Cs\'aki-et-al duality of Section \ref{sec:Slike}. In order to form ${}^C\CTb$, we are instructed to take a copy of $\CTb$, with index $\CI_\CT(\hat z;p,q)$, and to flip the baryons $u^au^b$ and anti-baryons $v^av^b$. The flips introduce new chiral fields $b_{ab}$ and $b'_{ab}$ of R-charge 1 and conjugate flavor charges. Therefore, the index becomes
\be \label{IC}
\CI^C_1(\hat z;p,q) = \prod_{{1\leq a<b\leq 4,\atop 5 \leq a<b\leq 8}} \Gamma\Big(\frac{\sqrt{pq}}{\hat z_a\hat z_b};p,q\Big)\times \CI_\CT(\hat z;p,q)\,.
\ee
This function only has manifest $W(SU(4)_L\times SU(4)_R)=S_4\times S_4$ symmetry.

We expect the index \eqref{IC} to equal $\CI_\CT(z;p,q)$ given an appropriate map of fugacities $\hat z$ and $z$. Recall that the meson operators of $\CTb$ and ${}^C\CTb$ must match, $q^aq^{b+4}=u^av^b$ ($a,b=1,...,4$) and the similarly the baryon and anti-baryon operators are identified as $q^aq^b = \epsilon^{abcd}b_{cd}$ and $q^{a+4}q^{b+4}=\epsilon^{abcd}b'_{cd}$ ($a,b,c,d=1,...,4$). Therefore, the fugacities should obey
\be z_az_{b+4} = \hat z_a\hat z_{b+4}\,,\quad z_az_b = \epsilon_{abcd}\frac{1}{\hat z_c\hat z_d}\,,\quad z_{a+4}z_{b+4} = \epsilon_{abcd}\frac{1}{\hat z_{c+4}\hat z_{d+4}}\,,
\notag \ee
where $a,b,c,d$ run from $1$ to $4$.
This fixes the map of fugacities to take the form
\be \label{Cfug}
 \hat z_a = \frac{z_a}{\alpha}\,,\qquad \hat z_{a+4}=\alpha\, z_{a+4}\,\qquad (a=1,...,4)\,,\ee
with $\alpha = (z_1z_2z_3z_4)^{\frac12}=(z_5z_6z_7z_8)^{-\frac12}$. Given this identification, a fundamental identity \cite{Spir-thetaintegrals} for the function $\CI_\CT$ ensures that indeed $\CI_\CT^C(\hat z;p,q) = \CI_\CT(z;p,q)$.

For further illustration, consider the indices of the Seiberg and Intriligator-Pouliot duals of $\CTb$. To form the Seiberg dual, we start with a copy of $\CTb$ --- whose quarks transform in a conjugate representation of $SU(8)$ relative to the original theory --- and flip the mesons. Therefore,
\be \CI_\CT^S(\hat z;p,q) = \prod_{1\leq a,b\leq 4} \Gamma\Big(\frac{\sqrt{pq}}{\hat z_a\hat z_{b+4}};p,q\Big)\times \CI_\CT(\hat z;p,q)\,, \label{IS} \ee
with
\be \label{Sfug}
\hat z_a = \frac{\alpha}{z_a}\,,\qquad \hat z_{a+4}=\frac{1}{\alpha\, z_{a+4}}\qquad (a=1,...,4)\,.
\ee
For the Intriligator-Pouliot dual, we take a copy of $\CTb$ and flip everything, resulting in
\be \CI_\CT^{IP}(\hat z;p,q) = \prod_{1\leq i<j\leq 8} \Gamma\Big(\frac{\sqrt{pq}}{\hat z_i\hat z_j};p,q\Big)\times \CI_\CT(\hat z;p,q)\,, \label{IIP} \ee
with $\hat z_i = z_i^{-1},\,i=1,...,8$. One finds that \eqref{IS} and \eqref{IIP} both equal the original $\CI_\CT(z;p,q)$. Indeed, this can be proven via repeated applications of the fundamental identity $\CI_\CT^C(\hat z;p,q) = \CI_\CT(z;p,q)$.

Altogether, just as the 71 theories dual to $\CTb$ are constructed by splitting the eight doublets into $4+4$ and flipping mesons, baryons, or both, the corresponding indices are constructed by transforming appropriate flavor fugacities and adding elliptic-Gamma prefactors for the flipped quark bilinears. The transformations of flavor fugacities (such as \eqref{Cfug} and \eqref{Sfug}) correspond directly to Weyl reflections $\pi$ in the $E_7$ root lattice that map one embedded $SU(8)$ root system to another \cite{Rains-hypgeom}. This makes the $W(E_7)$ duality network of Section \ref{sec:E7net} very concrete. The indices of all the dual theories agree.

\subsection{Doubled index}
\label{sec:ind-double}

As expected, the indices of most duals of $\CTb$ do not enjoy a manifest $S_8$ group of symmetries. Of course, they are all secretly symmetric under the $S_8$ that permutes the original fugacities $z_i$. But, due to the elliptic-Gamma prefactors, they only have an $S_4\times S_4$ symmetry permuting the new fugacities $\hat z_i$. Therefore, the original $S_8$ does not combine with a new $S_8$ to form a complete $W(E_7)$ symmetry.

We fixed this in Section \ref{sec:doubled} by doubling the theory $\CTb$ to form a theory $\CTT$ that does, conjecturally, have enhanced $E_7$ symmetry at a point in its marginal parameter space. The enhancement can be seen very easily in the sphere index.

Since $\CTT$ is a superpotential deformation of a product $\CT\times \wt \CT$, the  index of $\CTT$ is simply the product of two $\CTb$ indices,
\be \label{I11} \CI_{I\!I}(z;p,q) = \CI_\CT(z;p,q)\, \CI_\CT(z^{-1};p,q)\,, \ee
where all the flavor fugacities have been inverted in the second copy, $z_i\to z_i^{-1}$, to reflect the fact that quarks in $\wt \CTb$ transform in the conjugate representation of $SU(8)$. It is not hard to see that this index must take exactly the same form after we simultaneously apply any of the duality transformations to both halves of the product. That is,
\be \CI_{I\!I}(z;p,q) = \CI_{I\!I}(\hat z;p,q) = \CI_\CT(\hat z;p,q)\CI_\CT(\hat z^{-1};p,q) \ee
for any transformed choice of fugacities $\hat z_i$ --- corresponding to a $4+4$ splitting of the quarks and a flip of mesons or baryons --- with no additional elliptic Gamma functions. Indeed, for every factor $\Gamma(\sqrt{pq}\hat z_i^{-1}\hat z_j^{-1};p,q)$ generated by dualizing $\CT$, there will be a factor $\Gamma(\sqrt{pq}\hat z_i\hat z_j;p,q)$ generated by dualizing $\wt \CT$, and the two exactly cancel. The cancellation reflects the fact, discussed in Section \ref{sec:doubled}, that upon simultaneously dualizing both $\CT$ and $\wt \CT$ all gauge-invariant chiral matter becomes massive and can be integrated out.

Therefore, the function $\CI_{I\!I}(z;p,q)$ is fully invariant under all 72 copies of $S_8$ embedded in $W(E_7)$, which combine to generate a complete $W(E_7)$ symmetry. By expanding $\CI_{I\!I}(z;p,q)$ as a series in $p$ and $q$, one the coefficients grouped into characters of $E_7$, which of course count various groups of operators in the theory.

\subsection{A 4d-5d half-index}
\label{sec:ind-half}

A more interesting way to construct an index with $W(E_7)$ symmetry is to consider the boundary condition $\CBT$ for a 5d theory of 28 hypermultiplets. The appropriate index in this case is a ``half-index.'' It generalizes the 3d-4d half-index constructions of \cite[Sec 5.2]{DGG-index}.

To begin, we recall that the sphere index of a 5d $\CN=1$ theory with $SU(2)_R$ R-symmetry is given by \cite{Minwalla-restrictions, KKL-5dindex}
\be \label{I5d}
 \CI^{5d}_\CT(z;p,q;\eta) = \Tr_{\CH(S^4)}(-1)^Fe^{-\beta H} p^{\tfrac R2+J_1+J_2}q^{\tfrac R2+J_1-J_2}z^e\eta^k\,.
\ee
where $R$ is the R-charge for a Cartan generator of $SU(2)_R$; $J_1$ and $J_2$ are the angular momenta of states corresponding to Cartan generators of a fixed $SU(2)_1\times SU(2)_2\simeq SO(4)$ subgroup of the $SO(5)$ rotations of $S^4$; $e$ is a vector of flavor fugacities, and $k$ is the instanton number of states on $S^4$. Only states with $H=E-2J_1-\frac32R=0$ contribute.

In general, the presence of instanton sectors makes this index quite nontrivial to evaluate. However, all we will need here is the index of a free hypermultiplet, all of whose states have instanton number zero. For a hypermultiplet with flavor charge $f$, the index is \cite{KKL-5dindex}
\be \label{Ihyper} \CI^{5d}_{\rm hyper}(z;p,q;\eta) = \frac{1}{(\sqrt{pq}z^f;p,q)_\infty (\sqrt{pq}z^{-f};p,q)_\infty}\,,\ee
with $(z;p,q)_\infty = \prod_{m,n\geq 0}(1-p^mq^nz)$ as before.
The two terms in the denominator come from bosonic modes of the two chirals $(X,Y)$ in the hypermultiplet.
Note that supersymmetry fixes the R-charge of these chirals to equal 1.

We want to give the 5d theory of hypermultiplets a BPS boundary condition, and there is a natural way to do this in the index setup. The $SU(2)_1\times SU(2)_2\simeq SO(4)$ rotations whose fugacities appear in \eqref{I5d} preserve an equator $S^3\subset S^4$. By cutting $S^4$ in half along this equator, we can couple the 5d theory on a half-sphere $D^4 \times \R$ to a 4d $\CN=1$ theory at the boundary $S^3\times \R$, with the $U(1)_R$ R-symmetry of the 4d theory embedded in the $SU(2)_R$ 5d R-symmetry. Such a coupling preserves half of the bulk supersymmetry. In particular, it preserves the supercharge $Q$ with respect to which the index \eqref{I5d} is constructed. Restricted to the boundary, this supercharge also defines a 4d sphere index \eqref{4dpq} for the boundary theory. Altogether, we can define a 5d \emph{half-index} on $D^4\times \R$ (or a partition function on $D^4\times S^1$) that gets contributions both from 5d bulk degrees of freedom and from the 4d degrees of freedom at the boundary.

Now, consider the boundary setup of Section \ref{sec:bdy}. Suppose that a 5d bulk hypermultiplet is split into chirals $(X,Y)$, with charges $(+1,-1)$ under some $U(1)$ flavor symmetry. If we impose Neumann b.c. for $X$ and Dirichlet b.c. for $Y$ at the boundary of $D^4\times \R$, then the hyper will contribute
\be \CI^{\rm half}_{\pd_nX=0}(z;p,q) = \frac{1}{(\sqrt{pq}\,z;p,q)_\infty} \ee
to a half-index, since only modes of $X$ can be excited. Conversely, if we impose Dirichlet b.c. for $X$ and Neumann for $Y$, the half-index contribution is
\be \CI^{\rm half}_{\pd_nY=0}(z;p,q) = \frac{1}{(\sqrt{pq}\,z^{-1};p,q)_\infty}\,.\ee

This immediately allows us to demonstrate the action of a flip. We argued in Section \ref{sec:bdy} that giving Neumann b.c. to $Y$ (Dirichlet b.c. to $X$) and coupling it to chiral $\phi$ at the boundary through a superpotential $W=Y\phi$ was equivalent to giving Dirichlet b.c. to $Y$ and Neumann b.c. to $X$. In terms of half-indices, this equivalence is expressed by the equality
\be \frac{1}{(\sqrt{pq}z^{-1};p,q)_\infty}\Gamma(\sqrt{pq}z;p,q) = \frac{1}{(\sqrt{pq}z;p,q)_\infty}\,.\ee

We would like to construct the half-index of the boundary condition $\CBT$ and see that it has manifest $W(E_7)$ symmetry. To do so, we couple $\CTb$ to a 5d theory of 28 hypermultiplets $(X_{ij},Y^{ij})$ via the $SU(8)$-invariant superpotential $W = cM^{ij}X_{ij}$. This imposes a modified Dirichlet b.c. for $Y^{ij}$. Therefore, the 5d half-index is
\be \label{Ihalf}
 \CI^{\rm half}_{\CBT}(z;p,q)= \prod_{1\leq i<j\leq 8} \frac{1}{(\sqrt{pq}z_i^{-1}z_j^{-1};p,q)_\infty} \times \CI_\CT(z;p,q)\,.
\ee
Note that the fugacities $z_i^{-1}z_j^{-1}$ of $X_{ij}$ appearing in the denominator are inverse to the fugacities $z_iz_j$ of the quark bilinears $M^{ij}$. This index has manifest $S_8$ symmetry.

If we dualize the theory $\CTb$, flipping any combination of mesons and/or baryons, we will recover an index of the same form \eqref{Ihalf}, where the $z_i$ have been replaced by the dualized fugacities $\hat z_i$. For example, suppose that we apply a Seiberg duality and flip the mesons. Then
\begin{align} \CI^{\rm half}_{\CBT}(z;p,q) &=  \prod_{1\leq i<j\leq 8} \frac{1}{(\sqrt{pq}z_i^{-1}z_j^{-1};p,q)_\infty}  \prod_{1\leq a,b\leq 4} \Gamma\Big(\frac{\sqrt{pq}}{\hat z_a\hat z_{b+4}};p,q\Big)\times \CI_\CT(\hat z;p,q) \notag \\
&= \prod_{1\leq a<b\leq 4 \atop
 5\leq a < b \leq 8} \frac{1}{(\sqrt{pq}\hat z_a^{-1}\hat z_b^{-1};p,q)_\infty}\prod_{1\leq a,b\leq 4}\frac{1}{(\sqrt{pq}\hat z_a\hat z_b;p,q)_\infty} \notag  \\
 &\hspace{2in} \times \prod_{1\leq a,b\leq 4} \Gamma\Big(\frac{\sqrt{pq}}{\hat z_a\hat z_{b+4}};p,q\Big)\times \CI_\CT(\hat z;p,q) \notag \\
 &= \prod_{1\leq i<j\leq 8} \frac{1}{(\sqrt{pq}\hat z_i^{-1}\hat z_j^{-1};p,q)_\infty}  \times \CI_\CT(\hat z;p,q)  \notag \\
 &= \CI^{\rm half}_{\CBT}(\hat z;p,q)\,,
\end{align}
where in the second step we used the map of fugacities \eqref{Sfug} to rewrite the 5d chiral contribution. As discussed in Section \ref{sec:bdy}, we have arrived at the index of an equivalent boundary condition where a different half of the 5d bulk chiral multiplets are coupled to 4d boundary operators.

Altogether, it is easy to see that any transformation of fugacities $z_i\to \hat z_i$ is an exact symmetry of $\CI^{\rm half}_{\CBT}(z;p,q)$. Therefore, just like the doubled index $\CI_{I\!I}(z;p,q)$, it is simultaneously invariant under 72 $S_8$ groups embedded in $W(E_7)$, which combine to generate $W(E_7)$ itself.

It is amusing to verify that the expansion of $\CI^{\rm half}_{\CBT}(z;p,q)$ in $p$ and $q$ indeed contains characters of $E_7$. We find that the coefficient of $p^{\frac12} q^{\frac12}$ is the character of the fundamental representation, expressed as $\sum_{i<j} (z_iz_j+z_i^{-1}z_j^{-1})$, which clearly counts operators $M^{ij}$ and $X_{ij}$; while the coefficient of $pq$ is the character of the 1463-dimensional representation, and counts the marginal operators discussed in Section \ref{sec:bdy}.

\section{One dimension down, from $E_7$ to $SO(12)$}

Finally, we arrive at the initial motivation for this note: showing that a certain boundary condition for a 4d theory of 32 $\CN=2$ half-hypermultiplets has a hidden $SO(12)$ symmetry.

Our strategy shall be to start with the 4d boundary condition $\CBT$, with $E_7$ global symmetry, and to reduce it to a 3d boundary condition by compactifying on a circle. The group $E_7$ has $SO(12)\times SU(2)_m$ as a maximal subgroup, and by adding an $SU(2)_m$ flavor Wilson line around the compactification circle (as suggested by \cite{TV-6j}), we can break $E_7$ to $SO(12)$. This defines for us the 3d boundary condition $\CBT^*$. In the bulk, the theory $\CBT$ consisted of 56 half-hypermultiplets in the fundamental representation of $E_7$, which decompose as
\be 56 \;\to\; (32,1)+(12,2)\,, \ee
\ie\ a chiral spinor of $SO(12)$ (which is still a pseudo-real representation) and a bifundamental of $SO(12)\times SU(2)_m$. The bifundamentals are made massive by the $SU(2)_m$ Wilson line and can be integrated out, so we find that $\CBT^*$ is a boundary condition for the $32$ half-hypers.

Just as the 4d boundary condition $\CBT$ had multiple dual UV descriptions, related by flips and Seiberg duality, the boundary condition $\CBT^*$ acquires multiple descriptions related by flips and 3d mirror symmetry. With some care, these UV descriptions can be derived by compactifying the purely 4d theory $\CTb$ on a circle with a flavor Wilson line, to obtain a 3d $\CN=2$ theory $\CTb^*$ that has the appropriate chiral operators to couple to 32 bulk half-hypermultiplets.

Each of the 72 dual descriptions of $\CTb$, parametrized by a different embedding of an $SU(8)$ flavor group inside $E_7$, reduces to a dual description of $\CTb^*$. The 3d duals fall into two distinct classes, depending on how the 4d $SU(8)$'s intersect $SO(12)\times SU(2)_m$ in $E_7$.
One natural option is to intersect along $SU(6)\times SU(2)_m\times U(1)$, which --- after making the $SU(2)_m$ flavor symmetry massive --- leads to 3d $SU(2)$ gauge theory with six chiral doublets of quarks, or $N_f=3$ SQCD. We'll call this theory $\CTb^*_{(6)}$. 

The less obvious alternative is to intersect along $SU(4)\times SU(4)\times U(1)_m$. Reducing to 3d and turning on a real mass for the $U(1)_m$ vector symmetry naively makes all the quark doublets massive. One quark in each doublet can be kept light by turning on a large background value for the scalar in the $SU(2)$ gauge multiplet, which breaks the gauge symmetry down to $U(1)$. This leads to a description of $\CTb^*$ as $N_f=4$ SQED, with four chirals of charge $+1$ and four chirals of charge $-1$.
An analysis of twisted instantons in the circle compactification shows that the monopole and anti-monopole operators of 3d SQED must get added to its superpotential \cite{AHISS}. We call the resulting theory $\CTb^*_{(4,4)}$. 

Altogether, the 72 duals of $\CTb$ reduce to 32 descriptions of $\CTb^*$ as $SU(2)$ gauge theory with six chiral doublets, labelled by elements of the coset $W(SO(12))/W(SU(6))$; and 40 descriptions of $\CTb^*$ as $N_f=4$ SQED, labelled by elements of the coset $W(SO(12))/W(SU(4)\times SU(4))$. These dual theories can carefully be coupled to bulk hypermultiplets to form the $SO(12)$-invariant boundary condition $\CBT^*$, or doubled up to form a purely 3d $SO(12)$-invariant theory $\CTT^*$, the reduction of $\CTT$.
In the remainder of this section, we will briefly describe the couplings of $\CTb^*_{(6)}$ and $\CTb^*_{(4,4)}$ to bulk hypermultiplets, verify that marginal operators of $\CBT^*$ organize themselves into $SO(12)$ representations, and construct $SO(12)$-invariant partition functions. We hope that these details will form a more complete picture of the 3d setup. However, we emphasize that properties like the existence of enhanced flavor symmetry for $\CBT^*$ and $\CTT^*$ are simply ensured by the proper reduction from four dimensions.

\subsection{Couplings, flips, and operators}
\label{sec:3dcouple}

Consider the 3d theory $\CT^*_{(6)}$ in a little more detail. It has a manifest $U(6)=SU(6)\times U(1)_A$ flavor symmetry, and has fifteen gauge-invariant baryon operators $M^{ij}=q^iq^j$ in the $(15)_2$ representation of $U(6)$. (Here we denote the six fundamental quark doublets as $q^i$, $i=1,...,6$.) In addition, the theory has a monopole operator $\eta$ with axial charge $-6$. When coupling to sixteen 4d bulk hypermultiplets, both the baryon and monopole operators must be used. Indeed, a chiral spinor of $SO(12)$ (containing bulk half-hypers) decomposes under $SU(6)\times U(1)_A$ as
\be 32 \;\to \; (15)_{2}+(\ol{15})_{-2}+(1)_{-6}+(1)_{6}\,, \label{32U6} \ee
which is exactly right to match the baryon and monopole charges.

Splitting the 16 hypers as $(X_{ij},Y^{ij})_{i<j}$ and $(X,Y)$, a general $U(6)$-invariant coupling takes the form
\be W = c \sum_{i<j} X_{ij}M^{ij} + b X\eta\,. \label{Wbc} \ee
The couplings $b$ and $c$, however, are not independently marginal. A quick Z-extremization calculation \cite{Jafferis-Zmin} indicates that the IR R-charge assignment for $\CT^*_{(6)}$ has $R(q^i)\approx 0.37$ and $R(\eta) = 2N_f(1-R(q^i))-2=1.78$. After adding either one of the terms to \eqref{Wbc}, the theory flows to a new IR fixed point where $R(q^i)=1/2$ and $R(\eta)=1$, allowing for a single marginal deformation that preserves $U(6)$. (The dimension and R-charge of $X_{ij}$ and $X$ are fixed to 1 by $\CN=2$ supersymmetry in four dimensions; so if $R(M^{ij})=R(\eta)=1$ the couplings above become marginal.) This is similar to the situation in one dimension up, where \eqref{MX} admitted a single marginal deformation. Somewhere along the $U(6)$-invariant line, we must find the point $p^*$ where $\CBT^*$ has enhanced $SO(12)$ flavor symmetry.

Different mirror-symmetric descriptions of $\CT^*_{(6)}$ are systematically deduced from the Seiberg dualitites of $\CT$. The basic idea is to take a copy of $\CT^*_{(6)}$, split the quarks into two sets, and flip mesons or baryons. In order to stay within the family of $\CT^*_{(6)}$-like theories, it is necessary to take a splitting into $4+2$ quark doublets, breaking the flavor symmetry to $SU(4)\times SU(2)\times U(1)\times U(1)$, and to flip either $8$ ``mesons'' or $7$ ``baryons'' plus the monopole operator. There are $2\times {6\choose 2} = 30$ ways to do this. Together with the original theory and a fully flipped theory, this accounts for 32 different duals of $\CT^*_{(6)}$, labelled by embeddings of $U(6)$ in $SO(12)$.

In the presence of the bulk coupling \eqref{Wbc}, every flip gets ``absorbed'' by the bulk hypermultiplets, as described in Sections \ref{sec:doubled}--\ref{sec:bdy}. Thus, the dualities lead to different subsets of the bulk hypers coupled to $\CT^*_{(6)}$ itself, with different manifest $U(6)$ symmetries.
By following how the coupling \eqref{Wbc} transforms under mirror symmetry, one can argue directly that a point $p^*$ exists where multiple $U(6)$ symmetries are realized simultaneously, and produce the enlarged flavor group $SO(12)$ for $\CBT^*$.

In a similar fashion, we may describe the theory $\CT^*_{(4,4)}$ ($N_f=4$ SQED) and its duals. This theory has eight chirals $\phi^a,\,\wt\phi^a$, $a=1,...,4$, with $U(1)$ gauge charges $+1$ and $-1$, respectively. Since the monopole and anti-monopole have been added to the superpotential, the axial and topological $U(1)$ symmetries are broken, leaving a flavor symmetry group $SU(4)\times SU(4)$. There are sixteen gauge-invariant meson operators $M^{ab}=\phi^a\wt\phi^b$ that can couple to bulk hypermultiplets, which agrees well with the decomposition of the chiral spinor of $SO(12)$ into $SU(4)\times SU(4)$ representations,
\be 32\;\to\; (4,\ol 4)+(\ol 4,4)\,.\ee 
Now we should represent the bulk hypermultiplets as $(X_{ab},Y^{ab})$, and find a single $SU(4)\times SU(4)$-invariant coupling
\be W = c \sum_{a,b} X_{ab} M^{ab}\,. \label{Wphi} \ee
The presence of monopole operators in the superpotential fixes the R-charge of all the chirals to be $R(\phi^a)=R(\wt\phi^a)=1/2$, so this coupling is marginal.

In order to obtain mirror-symmetric descriptions of $\CT^*_{(4,4)}$, we can either flip all 16 meson operators, or split the $4+\ol 4$ chirals into $2+\ol 2+2+\ol 2$ (splitting both the $\phi^a$ and $\wt \phi^b$ into $2+\ol 2$) and flip eight gauge-invariant ``baryons'' or eight ``mesons'' with respect to the splitting. Generically, $SU(2)^4 \times U(1)^2$ flavor symmetry is preserved. This turns out to produce only 20 distinct theories, rather than the expected 40. After the reduction to three dimensions, the 40 duality frames of $\CT$ for which $SU(8)$ intersects $SO(12)$ as $SU(4)\times SU(4)$ collapse in pairs to these 20 theories. The direct argument for an $SO(12)$-invariant point $p^*$ for $\CBT^*$ follows without any major complications.

For completeness, we note that by taking a copy of $\CT^*_{(6)}$, splitting the quarks into $3+\ol 3$, and flipping 9 mesons (plus monopole) or 6 baryons, one obtains $\CT^*_{(6)}$-like duals to the 20 $\CT^*_{(4,4)}$-like theories. Conversely, by taking a copy of $\CT^*_{(4)}$, splitting the chirals into $3+\ol 1+\ol 3+1$, and flipping 10 mesons or 6 baryons, one finds $\CT^*_{(4,4)}$-like duals to the 32 $\CT^*_{(6)}$-like theories. In four dimensions, these transformations change the way $SU(8)$ intersects $SO(12)$.

Finally, let us confirm $SO(12)$ enhanced symmetry in $\CBT^*$ by performing a simple, direct count of exactly marginal operators. Take the $N_f=4$ SQED theory $\CT^*_{(4,4)}$ coupled via the superpotential \eqref{Wphi} to 16 bulk chiral multiplets $X_{ab}$ that have Neumann boundary conditions. In the decoupling limit, the flavor symmetry is $Sp(32)\times SU(4)\times SU(4)$. Along the line parametrized by the coupling \eqref{Wphi}, the flavor symmetry is broken to a diagonally embedded $SU(4)\times SU(4)$, and it is enhanced back to $SO(12)$ at $p^*$. Correspondingly, one finds 100 independent marginal operators of the form $M^{ab}M^{cd}=\phi^a\phi^c\wt \phi^b\wt \phi^d$, $16^2=256$ marginal operators of the form $M^{ab}X_{cd}$, and $\frac12 16(16+1)=136$ marginal operators of the form $X_{ab}X_{cd}$. In the last set, 36 of the $XX$ operators combine with broken \emph{bulk} currents of $SO(12)$ along the $SU(4)\times SU(4)$-invariant line, but are not lifted. On the other hand, 30 of the $MX$ operators are lifted by combining with broken off-diagonal currents of the $SU(4)\times SU(4)$ symmetries in the boundary and in the bulk. Therefore, $100+256+136-30=462$ exactly marginal operators remain, and fit into an irreducible representation of $SO(12)$. 
The marginal operators could also be associated with the $\dim Sp(56)-\dim SO(12)=462$ currents that are broken overall in the bulk by the coupling \eqref{Wphi}.

Note that this is not the counting that would have resulted from naively decomposing the 1463-dimensional representation of $E_7$, which held all the marginal operators of the 4d-5d boundary theory $\CBT$, under $SO(12)\times SU(2)_m$. Decomposing and keeping only $SU(2)$ singlets, we find
\be 1463\;\to\; (462,1)+(66,1)+\cancel{(\ol{352},2)}+\cancel{(77,3)}\,, \ee
with an extra adjoint $(66)$ of $SO(12)$. This adjoint contains operators that appear to be invariant under $SU(2)_m$, but whose constituent quarks actually become massive upon turning on the $SU(2)_m$ Wilson line.

\subsection{$SO(12)$-invariant partition functions}

The authors of \cite{TV-6j} constructed 3d ellipsoid $(S^3_b)$ partition functions for the 3d theories $\CT^*_{(6)}$ and $\CT^*_{(4,4)}$ that we introduced in this section. By comparing partition functions of these theories, with appropriate flips included, they verified that the theories and their multiple duals were related by 3d $\CN=2$ mirror symmetry. As long as one ignores extra flips, the partition functions in various duality frames  display either a discrete $W(U(6))\simeq S_6$ or $W(SU(4)\times SU(4))\simeq S_4\times S_4$ Weyl group symmetry, acting on flavor fugacities.

We are now in a position to produce 3d (or 3d-4d) partition functions that are fully invariant under the action of the Weyl group $W(SO(12))$, and manifestly invariant under either $S_6$ or $S_4\times S_4$ in a given duality frame. The construction is entirely analogous to that of 4d and 4d-5d indices in Section \ref{sec:indices}. Indeed, the $W(SO(12))$-invariant ellipsoid partition functions are direct analytic reductions of the $W(E_7)$-invariant indices in one higher dimension \cite{DSV-reduce, GY-reduce}.

Let us begin with the purely 3d theories $\CT^*_{(6)}$ and $\CT^*_{(4,4)}$. The ellipsoid $(S^3_b)$ partition function of an $\CN=2$ theory was defined in \cite{Kapustin-3dloc,HHL}. It depends on real masses $\mu_i$ corresponding to a maximal abelian subgroup of the flavor symmetry group, on the UV R-charge assignment used in coupling the theory to the curvature of $S^3_b$, and on the real parameter $b$. To obtain the partition function from a 4d index, one sets
\be p = e^{2\pi i \beta b}\,,\qquad q=e^{2\pi i\beta b^{-1}}\,,\qquad z_i=e^{2\pi \beta \mu_i}\,, \label{zmu} \ee  
where $z_i$ are flavor fugacities, and takes the (regularized) limit $\beta\to 0$. Then, for example, the commonly occurring products $(z;p,q)_\infty =\prod_{m,n\geq 0}(1-p^m q^nz)$ reduce to
\be (z;p,q)_\infty \overset{\beta\to 0}{\to} \prod_{m,n=0}^\infty (2\pi i \beta)(-i\mu+mb+nb^{-1}) = \frac{1}{\Gamma_b(-i\mu)}\,, \label{Gb} \ee
where the function $\Gamma_b(-i\mu)$ is a regularized version of the infinite product, with zeroes at $i\mu = mb+nb^{-1}$, $m,n\geq 0$ (\cf\ \cite{Barnes-QDL, DGOT, TV-6j}). The full partition function of a chiral multiplet with R-charge $r$ and real mass $\mu$ is
\be \Gamma\big((pq)^{\frac r2}z;p,q\big) \;\to\; \frac{\Gamma_b\big(\frac{Q}{2}r-i\mu\big)}{\Gamma_b\big(Q-\frac{Q}{2}r+i\mu\big)} \equiv S_b\big(\tfrac{Q}{2}r-i\mu\big)\,, \ee
with $Q=b+b^{-1}$. The function $S_b\big(\tfrac{Q}{2}r-i\mu\big)$ here was denoted $s_b\big(\tfrac{iQ}{2}-\tfrac{iQ}{2}r-\mu\big)$ in \cite{HHL}.

Using this dictionary, together with the additional fact that%
\footnote{These asymptotics reflect the fact that a very massive chiral can be integrated out to leave behind a pure Chern-Simons level for the $U(1)$ flavor symmetry that rotates its phase.} %
$S_b\big(\tfrac{Q}{2}-i\tau\big)\to e^{\mp \frac{i\pi}{2}\tau^2+{\rm const}}$ as $\tau\to\pm\infty$,
it is straightforward to derive the partition functions of $\CT^*_{(6)}$ and $\CT^*_{(4,4)}$ from four dimensions. One finds \cite{TV-6j} that
\be \CZ_b^{(6)}(\mu_i) = \frac12\int_\R d\sigma\, \frac{\prod_{i=1}^6S_b\big(\frac{Q}{4}-i\mu_i-i\sigma \big)S_b\big(\frac{Q}{4}-i\mu_i+i\sigma\big)}{S_b(2i\sigma)S_b(-2i\sigma)}\,, \label{Z6} \ee
with no constraint on the real masses $\mu_i$, and that
\be \CZ^{(4,4)}_b(\mu_1,...,\mu_4,\nu_1,...,\nu_4) = \int_\R d\sigma\,\prod_{a=1}^4 S_b\big(\tfrac Q4-i\mu_a-i\sigma\big)S_b\big(\tfrac Q4-i\nu_a+i\sigma\big)\,, \label{Z4} \ee
with $\mu_1+\mu_2+\mu_3+\mu_4=\nu_1+\nu_2+\nu_3+\nu_4=0$. These partition functions can be dualized and related in various ways using the flips described in Section \ref{sec:3dcouple}.

For example, the partition functions \eqref{Z6}--\eqref{Z4} are related to each other by
\be \CZ_b^{(4,4)}(\mu_a,\nu_b) = \prod_{a=1}^3 S_b\big(\tfrac Q2-i(\mu_a+\nu_4)\big)S_b\big(\tfrac Q2-i(\nu_a+\mu_4)\big)\times \CZ_b^{(6)}(\hat \mu_i)\,, \label{Z43} \ee
where $\hat \mu_i = \mu_i-\alpha,\, \hat \mu_{i+3}= \nu_i-\alpha$\; ($i=1,2,3$) with $\alpha = \frac12 (\mu_1+\mu_2+\mu_3+\nu_4)=\frac12(\mu_1+\mu_2+\mu_3-\nu_1-\nu_2-\nu_3)$. The six extra chirals in the prefactor on the RHS, all with R-charge 1, come from flipping the six ``baryons'' in a $3+\ol{3}$ split of the six quark doublets of $\CT^*_{(6)}$. Further identities are discussed in \cite{TV-6j}.
 
To form an $SO(12)$-invariant partition function, we should couple the 3d theory $\CTb^*$ (in some duality frame) either to 4d bulk hypermultiplets, forming the boundary condition $\CBT^*$, or to another copy $\wt \CTb^*$ with conjugate flavor symmetry, forming the doubled theory $\CTT^*$. In the case of a boundary condition, we obtain a ``half--$S^4_b$'' partition function, which contains the contributions from the bulk chirals that have Neumann b.c. as well as the $S^3_b$ partition function of the boundary theory. It is
\begin{align} \CZ_b^{\rm half}[\CBT^*](\mu_a,\nu_b) \label{Zhalf}
 &= \prod_{a,b=1}^4 \Gamma_b\big(\tfrac Q2+i(\mu_a+\nu_b)\big) \times \CZ_b^{(4,4)}(\mu_a,\nu_b) \\
 &= \Gamma_b\big(\tfrac Q2-i(\hat\mu_1+...+\hat\mu_6)\big) \prod_{1\leq i<j\leq 6} \Gamma_b\big(\tfrac Q2+i(\hat\mu_i+\hat\mu_j)\big)\times \CZ_b^{(6)}(\hat \mu_i)\,, \notag
\end{align}
where in the first line we have used a coupling to $\CT^*_{(4,4)}$ and in the second we used the dual coupling to $\CT^*_{(6)}$, and the fugacities are related as below \eqref{Z43}. One can explicitly see the contributions $\Gamma_b\big(\frac Q2-...\big)$ from the chiral halves of 16 bulk hypermultiplets, with charges matching 16 boundary mesons in one case, and 15 baryons plus a monopole in the other. For comparison, note that the full $S^4_b$ partition function of a hypermultiplet with mass $\mu$ is \cite{Pestun-S4, AGT, HH-S4b}
\be \CZ_b^{\text{4d hyper}}(\mu) = \Gamma_b\big(\tfrac Q2-i\mu)\,\Gamma_b\big(\tfrac Q2+i\mu)\,, \ee
and the two factors here are naturally associated to the chirals $(X,Y)$ in the hypermultiplet.

The half--$S^4_b$ partition function \eqref{Zhalf} is invariant under flips, by construction. Thus, it possesses all 32 different $S_6$ symmetries and all 20 different $S_4\times S_4$ symmetries, embedded in the full symmetry group $W(SO(12))$.

We can easily write down a $W(SO(12))$-invariant ellipsoid partition function for the doubled theory $\CTT^*$ as well. It takes the form
\be \CZ_b^{I\!I}(\mu_a,\nu_b) =  \CZ_b^{(4,4)}(\mu_a,\nu_b)\CZ_b^{N_f=4}(-\mu_a,-\nu_b) = \CZ_b^{(6)}(\hat \mu_i)\CZ_b^{(6)}(-\hat \mu_i)\,, \ee
with the usual relation between masses $\mu_a,\nu_b$ and $\hat \mu_i$.

Finally, it is amusing to consider 3d indices, or 3d-4d half-indices \cite{KMMR-index, IY-index, KW-index, DGG-index} that enjoy full $W(SO(12))$ invariance. With conventions as in \cite{DGG-index}, the half-index of the  boundary condition $\CBT^*$ is
\begin{align}
\CI^{\rm half}[\CBT^*](\zeta_a,\xi_b;q) &=
 \prod_{a,b=1}^4\frac{1}{\big(q^{\frac12}\zeta_a^{-1}\xi_b^{-1};q\big)_\infty} \\
 &\quad \times \sum_{m\in \Z} \oint \frac{d\sigma}{2\pi i\sigma} \sigma^{4m}  \prod_{a=1}^4 \CI_\Delta(q^{\frac14} \zeta_a\sigma,m;q)\,\CI_\Delta(q^{\frac14} \xi_a\sigma^{-1},-m;q) \notag \displaybreak[0] \\
 &= \frac{1}{\big(q^{\frac12}\hat\zeta_1\hat\zeta_2\hat\zeta_3\hat\zeta_4\hat\zeta_5\hat\zeta_6;q\big)_\infty}\prod_{1\leq i<j\leq 6} \frac{1}{\big(q^{\frac12}\hat\zeta_i^{-1}\hat \zeta_j^{-1}\big)_\infty}  \\
 &\quad \times \frac12 \sum_{m\in \Z}\oint \frac{d\sigma}{2\pi i\sigma} s^{2m} \frac{\prod_{i=1}^6 \CI_\Delta\big(q^{\frac14}\hat\zeta_i\sigma,m;q\big)\,\CI_\Delta\big(q^{\frac14}\hat\zeta_i\sigma^{-1},-m;q\big)}
 { \CI_\Delta(\sigma^2,2m;q)\,\CI_\Delta(\sigma^{-2},-2m;q)}\,, \notag
\end{align}
with fugacity constraints $\prod_{a=1}^4 \zeta_a=\prod_{a=1}^4\xi_a = 1$ and $\hat \zeta_i = \zeta_i/\alpha,\, \hat \zeta_{i+3}=\alpha \xi_i\; (i=1,2,3)$, $\alpha = \sqrt{\zeta_1\zeta_2\zeta_3\xi_4}$; as well as $\CI_\Delta(\zeta,m;q)=\big(q^{1-\frac m2}\zeta^{-1};q\big)_\infty\big/\big(q^{-\frac m2}\zeta;q\big)_\infty$ and $(x;q)_\infty =\prod_{n=0}^\infty (1-q^nx)$.
We have written both the manifest $S_6$ and $S_4\times S_4$ versions of this function.
By expanding in powers of $q$, we can recover the $SO(12)$-invariant operator content of the theory. For example, setting all flavor fugacities to 1, we find
\be \CI^{\rm half}[\CBT^*](\zeta_a=\xi_a=1;q) = 1+32q^{\frac12}+462 q+\ldots\,,\ee
confirming our count of 462 marginal operators in an irreducible representation of $SO(12)$.

\acknowledgments{We would like to thank S. Razamat, N. Seiberg, J. Teschner, and G. Vartanov for helpful and valuable discussions. T.D. is supported primarily by the Friends of the Institute for Advanced Study, and in part by DOE grant DE-FG02-90ER40542. The research of DG was supported by the Perimeter Institute for Theoretical
Physics.  Research at Perimeter Institute is supported by the
Government of Canada through Industry Canada and by the Province of
Ontario through the Ministry of Economic Development and Innovation.}


\bibliographystyle{JHEP_TD}
\bibliography{toolbox}

\end{document}